\documentclass[final,floatfix,twocolumn,prb,amssymb,amsfonts,amsmath,superscriptaddress,showpacs]{revtex4}

\usepackage{graphicx}

\begin{document}

\title{Field induced magnetic ordering transition in Kondo insulators}

\author{Igor Milat} \affiliation{Theoretische Physik,
  ETH-H\"onggerberg, CH-8093 Z\"urich, Switzerland}
\author{Fakher Assaad} \affiliation{Universit\"{a}t W\"{u}rzburg, Am
  Hubland, D-97074 W\"urzburg, Germany} \affiliation{Theoretische
  Physik, ETH-H\"onggerberg, CH-8093 Z\"urich, Switzerland}
\author{Manfred Sigrist} \affiliation{Theoretische Physik,
  ETH-H\"onggerberg, CH-8093 Z\"urich, Switzerland}

\begin{abstract}
  We study the 2D Kondo insulators in a uniform magnetic field using
  quantum Monte Carlo simulations of the particle-hole symmetric
  Kondo lattice model and a mean field analysis of the Periodic
  Anderson model. We find that the field induces a transition to an
  insulating, antiferromagnetically ordered phase with staggered moment
  in the plane perpendicular to the field. For fields in excess of the
  quasi-particle gap, corresponding to a metal in a simple band
  picture of the periodic Anderson model, we find that the metallic
  phase is unstable towards the spin density wave type ordering for
  any finite value of the interaction strength. This can be
  understood as a consequence of the perfect nesting of the particle
  and hole Fermi surfaces that emerge as the field closes the gap.
  We propose a phase diagram and investigate the quasi-particle and
  charge excitations in the magnetic field. We find good
  agreement between the mean-field and quantum Monte Carlo results.
\end{abstract}


\maketitle

\section{Introduction}
\label{sec:Introduction}

Kondo insulators, or heavy fermion semiconductors, are materials
containing at least one atom per formula unit with a partially filled
$f$ or $d$ shell and exhibiting properties similar to very narrow
gap semiconductors.  $CeRhAs$, $CeRhSb$, $YB_{12}$, $Ce_3Bi_4Pt_3$ and
$SmB_6$ are the most thoroughly investigated examples
\cite{riseborough.00}. In the canonical model, the formation of the
gap in Kondo insulators is a consequence of the hybridization between
the conduction band and the effective $f$-electron level which gives
rise to quasi-particle and spin-gaps at low temperatures.  Adopting a
band picture one can
close the gap by applying a high magnetic field, since the gap is on
the meV scale.  Although experiments on 
$YB_{12}$\cite{sugiyama.88}, $SmB_6$\cite{cooley.99} and
$Ce_3Bi_4Pt_3$\cite{jaime.00} seem to support this simple picture, the
exact nature of the field induced insulator to metal transition as
well as the role played by the strong correlations remains far from
understood.

Magnetic instabilities of the periodic Anderson model at half-filling
have been studied extensively by slave boson mean field
approximations\cite{riseborough.92,dorin.92,doradzinski.98}. A phase
diagram as a function of the interaction strength was established and
some thermodynamic and transport properties have been calculated. In
these studies, the only effect of the magnetic field is assumed to be
the stabilization of the ferromagnetically ordered state with respect
to other magnetic configurations.

Carruzzo and Yu\cite{carruzzo.1995} studied the one dimensional,
half-filled Kondo lattice in magnetic field using DMRG and bosonisation
techniques.  They found that although the spin gap closed at the
critical field, the charge gap remained due to umklapp scattering
They conclude that the 1D half-filled Kondo lattice is insulating at
all fields.

The effect of the field in disordered Kondo insulators was treated by
CPA in Ref.\onlinecite{deolivera.99,deolivera.03}. The authors find that, in the
absence of magnetic ordering, the magnetic field induces the insulator
to metal transition in the universality class of density driven
metal-insulator transitions. Based on scaling arguments the field
dependence of the quasi-particle gap as well as the critical field as
a function of temperature and impurity concentration were derived.

In this paper we present a detailed study of the field induced quantum
phase transition in 2D particle-hole symmetric models of Kondo
insulators.  We present a mean field calculation appropriate for the
small-$U$ limit of the periodic Anderson model. We find that the magnetic field induces a phase transition from the
paramagnetic insulator into a canted antiferromagnetic insulator which
remains stable at all field strengths (until all the electrons in the
system align with the field). While zero-energy spin modes exist, we find that
the field does not close the 
quasi-particle gap, if the lattice is bipartite, so that the metallic
ground state is never induced by the field. We investigate more
carefully the large-field limit using two effective models and reach
essentially the same conclusion - on the bipartite lattice, the
interaction is a relevant perturbation and the ground state remains
insulating at all fields. The approximate treatments are complemented
by a quantum Monte Carlo study of the particle hole symmetric Kondo
lattice model in 2D. We find good agreement between the results.

Recently, Beach and collaborators studied the effect of the magnetic
field on the Kondo insulators using a large-$N$ type mean field analysis
of the Kondo lattice model and quantum Monte Carlo simulations
\cite{beach.03}. They find that a large enough magnetic field induces a phase
transition to a metallic ground state from the insulating canted
antiferromagnetic state. The phase transition into the metallic state
occurs when the $f$ moments decouple from the conduction band, i.e. the
hybridization mean field vanishes at a certain critical field. The
question naturally arises whether this phase transition is real or
possibly an artifact of the large-$N$ mean field approach. 
Here we will show results which lead to a different
conclusion: the insulating state induced by the magnetic field remains
stable up to the full polarization of all electrons in the system, if
the system is completely particle-hole symmetric. 
Thus the particle-hole symmetric Kondo insulator is an insulator at
all fields. 

The paper is organized as follows:
In the next section we introduce the models used to describe Kondo
insulators.  In section~\ref{sec:meanfield} the phase diagram in the
presence of the magnetic field is obtained using a mean field
approximation for the half-filled periodic Anderson model. In
section~\ref{sec:largeB} we present a discussion of the Kondo lattice
model in high magnetic fields. In section~\ref{sec:qmc} the results of
the Quantum Monte Carlo simulations are presented and compared with
the mean field calculations. We summarize our results in
section~\ref{sec:conclusion} and briefly comment on their relevance
for the experimental systems.

\section{models}
\label{sec:models}

The canonical model used to describe the physics of the Kondo
insulators is the periodic Anderson model (PAM)\cite{riseborough.00}.
The PAM Hamiltonian, including the uniform magnetic field in the $z$-direction is
\begin{multline}
  \label{eq:Kpam}
  H_{PAM} = -\sum_{<i,j>,\sigma} t_{ij} c^\dag_{i\sigma} c_{j\sigma}
  + \epsilon_f \sum_{i \sigma} f^\dag_{i\sigma} f_{i\sigma}
  + U \sum_{i} n^f_{i\uparrow}n^f_{i\downarrow} \\
  + \sum_{k,\sigma} (V f^\dag_{k\sigma} c_{k\sigma} + \text{H.c.})  -
  g \mu_B \vec B \cdot \sum_i (\vec S^f_i + \vec S^c_i)
\end{multline}
Here all the symbols have their usual meaning. The PAM describes a two
band system in which one band (conduction electron, $c$ band) is
dispersive and uncorrelated and the other ($f$ band) dispersionless
and strongly correlated.  $t_{ij}$ is the hopping matrix element in
the $c$ band and $U$ the local Coulomb interaction in the $f$
band.
The two bands are mixed and the hybridization matrix element $V$
controls the mixing strength. In the particle-hole symmetric
model that we consider in the following, $t_{ij} = t$ for nearest
neighbor sites on the square lattice and zero otherwise and
$\epsilon_f = -U/2$. The magnetic field is coupled to the $c$ and $f$
electron spins only. The $g$ factors of the $c$ and $f$ electrons are
chosen to be the same, $g_c = g_f = 2$, for simplicity but choosing
them differently would not change the qualitative aspects of our
conclusions.  In the following, the magnetic field is measured in the
units of Zeeman energy.

In the non-interacting ($U=0$) case the ground state of the PAM is a
paramagnetic band insulator with the quasi-particle gap $\Delta_{qp}^0
= \sqrt{(W/2)^2 + V^2} - W/2 \simeq \frac{V^2}{W}$, where $W$ is the
conduction electron bandwidth.  In the field, the
Zeeman splitting reduces the quasi particle gap. For fields larger
than $B_{c1} = \Delta_{qp}^0$, the gap vanishes and the ground state
is metallic.  In the fields beyond $B_{c2} = \sqrt{(W/2)^2 + V^2} +
W/2 \simeq W + 2V^2/W$, all the spins are aligned with the field.  The
fully polarized ground state consists of two completely filled bands
and is a trivial band insulator.

Because of the particle-hole symmetry, the Fermi surfaces of the spin
up electrons and the spin down holes in the metallic state at
intermediate fields are perfectly nested with respect to $Q =
(\pi,\pi)$. The staggered susceptibility in the plane perpendicular to
the field diverges logarithmically as $\omega \rightarrow 0$. This
divergence makes the state unstable under perturbations coupling to
the staggered magnetization. In particular, one expects that a
staggered magnetization will be induced by any non-zero correlation on
the $f$ sites. The
ensuing ordered state is a canted antiferromagnet,
characterized by both $m_z$ and $m_x$ different from zero.

When $U$ is large enough ($U/V \gg 1$) to suppress charge
fluctuations on the $f$ sites, the low-energy physics of PAM is well
described by the Kondo lattice model
(KLM)\cite{tsunetsugu.1997,sinjukow.02},
\begin{equation}
  \label{eq:Kklm}
  H_{KLM} = -t \sum_{ \langle i,j \rangle  \sigma} c^\dag_{i\sigma}
  c_{j\sigma}  
  + J \sum_i \vec S^c_i \cdot \vec S^f_i - 2 B_z \cdot \sum_i (S^{z,f}_i
    + S^{z,c}_i)
\end{equation}
In the KLM, the charge fluctuations on the $f$ sites are completely
suppressed, $f$ electrons are treated as spins and the hybridization
is replaced by an antiferromagnetic exchange interaction between
conduction electrons and $f$ spins. Formally PAM and KLM can be
related by the Schrieffer-Wolff
transformation\cite{sw.66,sinjukow.02}, yielding $J = 8V^2/U$.

The zero-temperature, zero-field phase diagram of the $2D$
particle-hole symmetric KLM has been well established by various
numerical methods\cite{shi95,capponi.2000,zheng.03}.  In the absence
of the magnetic field the ground state of the KLM is a paramagnetic
insulator at large $J/t$. There is a quantum critical point at $J/t
\simeq 1.4$ and for small $J/t$ the ground state is
antiferromagnetically ordered.

The large $J/t$ paramagnetic state of the KLM is adiabatically
connected to the $U = 0$ state of the PAM. In the particle-hole
symmetric case, on finite lattices, this is guaranteed by theorems for
the ground states of the two models~\cite{tsunetsugu.1997}.

\section{Mean field analysis}
\label{sec:meanfield}

In this section we investigate the effect of the magnetic field on the
small-$U$ PAM. In particular we want to investigate the spin density wave
instability of the metallic state induced by the field in the
non-interacting model. To this end, we perform the mean field decoupling of
the interaction term in Eq.(\ref{eq:Kpam}) by assuming the
magnetization of the $f$ spins to have a uniform component along the
field axis and a staggered component in the plane perpendicular to the
field, $\langle \vec S^f_i \rangle = \vec m_i$ with $\vec m_i =
\left( (-)^i m_x, 0, m_z \right)$. This yields the mean field 
Hamiltonian (see the appendix~\ref{sec:app-meanfield} for details of
the derivation),
\begin{widetext}
\begin{multline}
  \label{eq:HMF}
  H_{MF} = \sum_{k,\sigma} (\epsilon_k - p_\sigma B) c^\dag_{k\sigma}
  c_{k\sigma} + \sum_{k,\sigma} (- p_\sigma )(B +U m_z)
  f^\dag_{k\sigma} f_{k\sigma} + V \sum_{k,\sigma} ( c^\dag_{k\sigma}
  f_{k\sigma} +
  f^\dag_{k\sigma} c_{k\sigma} ) \\
  - U m_x \sum_k ( f^\dag_{k+Q\uparrow} f_{k\downarrow} +
  f^\dag_{k+Q\downarrow} f_{k\uparrow} )  + N U (m_x^2 +
  m_z^2),
\end{multline}
with $p_\sigma = 1(\uparrow), -1(\downarrow)$ and
$\epsilon_k = -\tfrac W 2 [\cos(k_x) + \cos(k_y)]$.
%
%
The mean field Hamiltonian is quadratic in fermion operators and is easily
diagonalized by a unitary transformation. In the presence of the
staggered magnetization, the Brillouin zone is halved and one finds 8
quasi particle bands; the particle bands
\begin{multline}
  E_{p,\pm}^{\sigma}(k) = \frac 1 {\sqrt 2}
  \Bigl[ (B + U m_z)^2 + (U m_x)^2 + 2 V^2 + (\epsilon_k - p_\sigma B)^2 \pm \\
  \pm \sqrt{ ((U m_x)^2 + (B + Um_z)^2 - (\epsilon_k - p_\sigma B)^2
    )^2 + 4V^2[(\epsilon_k - 2 p_\sigma B - p_\sigma U m_z)^2 + (U
    m_x)^2] } \Bigr]^{1/2}
\end{multline}
and the hole bands related by, $E_{h,s}^\sigma(k) =
-E_{p, s}^{\bar \sigma}(k)$. Note that the $k$ dependence of the
quasi-particle bands originates only from the dispersion of the conduction
electrons.
On a bipartite lattice, with $\epsilon_{k+Q} =
-\epsilon_k$, the quasi particle bands satisfy, $E_{h,s}^\sigma(
\epsilon_k ) = -E_{p,\bar s}^{\bar \sigma} ( \epsilon_{k+Q} )$.
\end{widetext}

In the ground state, the particle bands are empty and the hole bands
are completely filled. To obtain the ground
state energy, the expression
\begin{equation}
  \label{eq:egs}
  E_{gs} = \sum_{s,\sigma}{\sum_k}' E_{h,s}^\sigma(k) 
  + \frac{(Um_x)^2 + (Um_z)^2}{U}
\end{equation}
must be minimized with respect to $m_x$ and $m_z$, yielding the usual
mean-field equations,
\begin{equation}
  \label{eq:mf}
  \frac{\partial E_{gs}}{\partial(U m_{x,z})} = 0.
\end{equation}
The prime on the
summation sign in equation~\ref{eq:egs} indicates that the summation
is to be taken over the magnetic Brillouin zone.

\subsection{Mean field phase diagram}
\label{sec:mean-field-phase}

\begin{figure}[tb]
  \centering
  \includegraphics[width=\columnwidth]{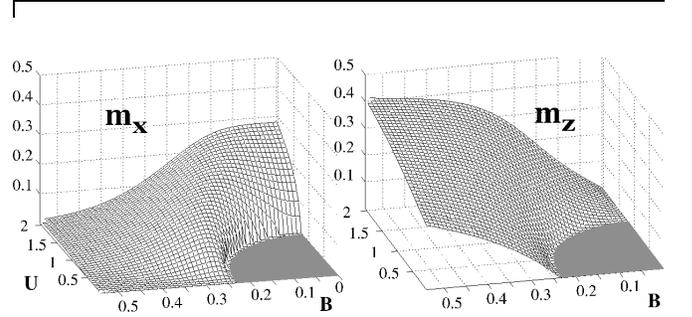}
  \caption{Staggered (left) and parallel magnetizations vs. B and U
    for $W=1, V=1$ obtained by numerically minimizing the mean field
    equations. The grayed out plateau marks the paramagnetic phase.}
  \label{fig:magnetizations}
\end{figure}

The minimization of the ground state energy for a range of $U$ and $B$
values was performed numerically and the obtained magnetization values
are shown in Fig.~\ref{fig:magnetizations}.
In zero field, the system is paramagnetic at small values of $U$ and
antiferromagnetically ordered beyond $U_c \simeq 1.25 V$. The
staggered magnetization grows as $(U-U_c)^{1/2}$ close to $U_c$ and
tends to the fully saturated value $m_x = 1/2$ as $U \rightarrow
\infty$.

The magnetic field applied to the system reduces the value of $U$ at
which the magnetic instability occurs. The phase boundary can be
obtained by solving
\begin{equation}
  \label{eq:Uc}
   \frac 2 {U_c(B)} = 
   -
  \left. \frac{\partial^2 E_0}{\partial (U m_x)^2}
  \right|_{m_x=0} ,
\end{equation}
where $E_0 = \sum_{s=\pm,\sigma=\uparrow\downarrow} \sum_k
E_{h,s}^\sigma (k)$.  At small fields, $B \ll B_{c1}$, the critical
interaction strength falls off as the square root of the field,
$U_c(B) - U_c(0) \propto -\sqrt{B}$, as expected in the mean field
approach.  At the phase boundary one finds the usual mean field
critical exponents for the staggered magnetization, $m^x_s \propto (U -
U_c(B))^{1/2}$ and $m^x_s \propto (B-B_c(U))^{1/2}$. After the
initial rise, $m_x$ goes through a maximum and falls of exponentially
in large fields.
On the phase boundary, the parallel susceptibility vanishes (it is
zero in the paramagnetic phase, since the spin excitations are gaped).
Close to the phase boundary, it behaves like
$m_z \propto (B-B_c(U))^\alpha$, with $\alpha > 1$.

The right hand side of the Eq.(\ref{eq:Uc}) is proportional to the static
staggered susceptibility of the $U m_x = 0$ state, $\chi_0^{+-}(Q)$. This can be expressed
using the familiar Lindhardt formula which in the case considered here
reduces to
\begin{equation}
  \label{eq:chi0}
  \chi_0^{+-}(Q) = - \left. {\sum_{k}}'
  \frac{f(E_{h,-}^\downarrow(k)) - f(E_{p,-}^\uparrow(k))}
  {E_{h,-}^\downarrow(k) - E_{p,-}^\uparrow(k)}\right|_{Um_x=0} .
\end{equation}
In small fields, the quasi-particle gap provides a cut-off for the
denominator in the sum on the right hand side and $\chi_0^{+-}(Q)$ is
finite. When the field closes the gap, the denominator vanishes along
the Fermi surface (Fermi lines), determined by the equation,
\begin{equation}
  \label{eq:epsilon0}
  \epsilon_k = \pm \epsilon_0 = \pm \frac{V^2- B U m_z - B^2}{B+U m_z}.
\end{equation}
Consequently, the staggered susceptibility diverges
logarithmically in the field $B > \Delta_{qp}^0$ and there is no
finite $U_c$. The system is
ordered for any finite interaction strength. The divergence of
$\chi_0^{+-}(Q)$ is a direct consequence of the perfect nesting,
$E_{h,-}^\downarrow(k) = - E_{p,-}^\uparrow(k)$, and is found at all fields, if
the conduction electron hopping is constrained to a bipartite lattice
and the system is half filled.\footnote{ Note that, the shape of the
  Fermi lines is determined only by the conduction electron dispersion
  and the surface they enclose by the half-filling condition.
  Therefore, the perfect nesting in the half-filled system
  can not be removed by changing the ratio of $g$ factors or by
  changing $\epsilon_f$.}

The behavior of the staggered magnetization at small $U$ can be found
by solving the mean field equations to leading logarithmic order in $U
m_x$. The details of the calculation are described in the
appendix~\ref{sec:app-meanfield}. The resulting expression for the
magnetization is
\begin{equation}
  \label{eq:mx}
  m_x \propto \exp\left[ - \frac{(B+Um_z)^4+ V^2(B+Um_z)^2}
    {V^4 \rho_0 U} \right],
\end{equation}
where $\rho_0$ is taken to be the density of states on the $m_x = 0$
Fermi surface.
It is interesting to note that the expression (\ref{eq:mx}) is valid for
the large field region and for the large $U$ region with $B \gg
8V^2/U$.

\subsection{Quasi-particle spectrum}
\label{sec:quasi-particle-gap}

\begin{figure}[tb]
  \centering
  \includegraphics[width=\columnwidth]{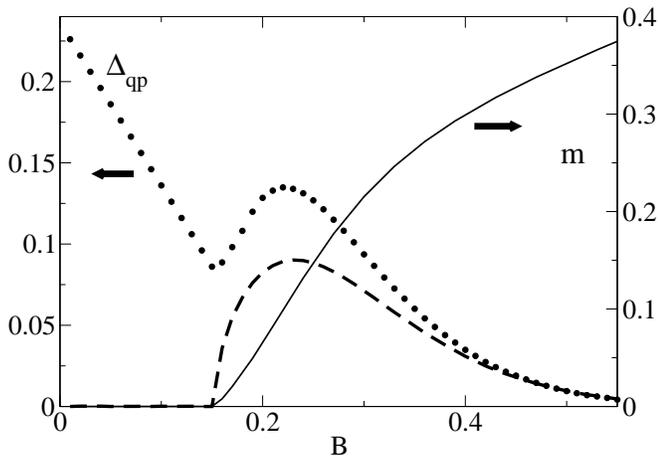}
  \caption{Quasi-particle gap (dots), staggered magnetization (dashed line)
  and total ($f$ + $c$) parallel magnetization (thin solid line)
  vs. magnetic field for $W = 4, V=1, U=1$. In the paramagnetic phase
  the gap decreases linearly
  with field. At large fields the gap follows the staggered
  magnetization.}
  \label{fig:qp-gap}
\end{figure}

The field dependence of the quasi-particle gap for a fixed value of the
interaction is shown in the Fig.~\ref{fig:qp-gap}.
In the paramagnetic phase, the gap decreases linearly with the
field. In the ordered phase, the quasi-particle gap
is proportional to the staggered moment and follows the same
exponential dependence for large fields. It is important to realize that
the quasi-particle gap always remains finite, so that the system is
insulating.

\begin{figure}[tb]
  \centering 
  \includegraphics[width=\columnwidth]{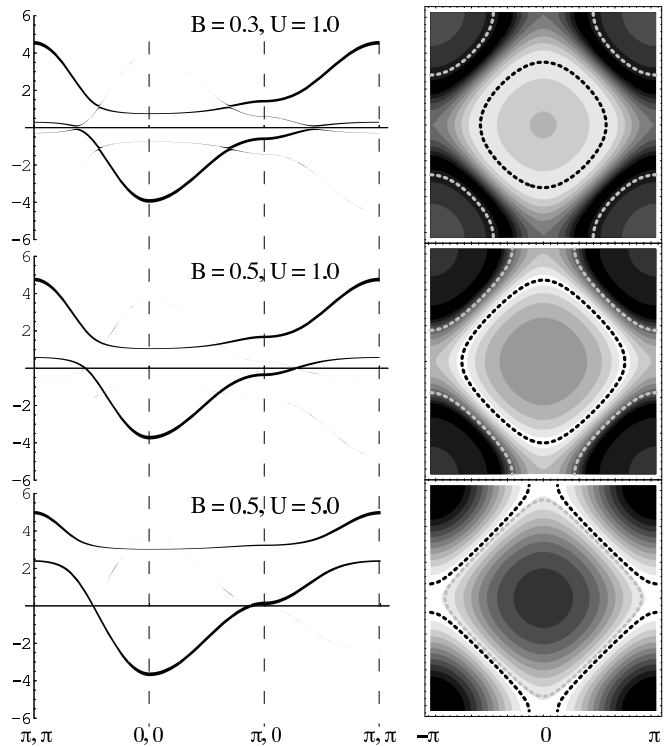} \\
  \caption{Poles of the $\sigma=\downarrow$ conduction electron
    spectral functions along the high symmetry lines of the Brillouin
    zone for field and
    interaction strengths indicated in the plots. The width of the
    lines indicates the weight in the pole.
    The plots on the right show the contours of constant
    quasi-particle gap magnitude with dashed
    lines indicating the position of the Fermi surfaces of the $m_x = 0$
    state.}
  \label{fig:bands}
\end{figure}

The spectral functions for the electrons in the mean field model
show infinitely sharp peaks at the quasi-particle band energies.
The poles of the $\sigma=\uparrow$ electron spectral function in the
ordered phase at various values of the 
field and interaction strengths are shown in Fig.~\ref{fig:bands}.
The width of the lines in the figure indicate the weights in
the corresponding poles.
Note that the gap at the Fermi surface is always finite, even though
the exponentially small scale is not immediately apparent in the plots.
The contour plots show the quasi-particle gap size in
the Brillouin zone. The gap minima are indicated by the dashed
lines in the contour plots.
The location of the gap minima indicates also the position of the Fermi
surfaces of the $m_x = 0$, metallic state.

It is interesting to observe the change in the character of the
quasi-particles at the gap minima as the field and the interaction
strength are varied.
At small $U$ and $B \simeq \Delta_{qp}^0$, the low-energy
quasi-particles are ``heavy'' and the
minimum of the gap lies near the zone center. As the field is
increased, the minimum moves towards the zone diagonal and the
quasi-particles become more and more $c$ like.

When the gap minimum reaches the zone diagonal, the
magnetization of the system along the field direction is exactly one half of
the fully saturated value. The field strength at which this happens,
$B_{1/2}$, depends on the interaction strength and can be obtained by
setting $\epsilon_0 = 0$ with the limiting behavior:
\begin{equation}
  \label{eq:5}
  B_{1/2} \rightarrow 
  \begin{cases}
    V, & U \rightarrow 0 \\
    \frac{2V^2}{U}, & U \rightarrow \infty
  \end{cases}.
\end{equation}
In the large $U$ limit, $B_{1/2}$ sets the energy scale at which
the $f$ electrons align with the field.

\section{Large B limit of the Kondo model}
\label{sec:largeB}

For large values of $U$, in the fields $B > B_{1/2}$, the $f$
electrons are almost completely
aligned with the field. At the mean field level, the poles of the
spectral function corresponding to the charge fluctuations on the $f$
sites move towards $\pm (B + U m_z)$, i.e. far from the Fermi
level.
In the large-$N$ mean field theory this eventually results in the
complete decoupling of the $f$ electrons from the $c$ band and the
decoupled metallic state obtains.\cite{beach.03}

In the particle hole symmetric case, the Fermi surfaces of the metallic
state are perfectly nested. The perfect
nesting makes the metallic state unstable at all fields in the small
$U$ limit of the PAM. We will now demonstrate that also in the limit
of large Coulomb interaction, i.e. for the KLM, the same instability
arises.

\subsection{Effective Hamiltonian approach}
\label{sec:effH}

We consider the KLM Hamiltonian in the large magnetic field $B \gg J$.
In the magnetic field, the $J=0$ ground state of the KLM is non
degenerate and is given by $| \psi \rangle = \prod_{k<k_{F\uparrow}}
c^\dag_{k\uparrow}\prod_{k<k_{F\downarrow}} c^\dag_{k\downarrow}
\prod_i f^\dag_{i\uparrow} |\rangle$. Flipping the $f$ spin is an
excitation with a gap given by the Zeeman energy.  A canonical
transformation approach can be used to generate an expansion in
$(J/B)$ around the $J=0$ ground state. The effective Hamiltonian
governing the low energy dynamics of the system is given by (the
details of the derivation in the appendix~\ref{sec:app-largeB}),
\begin{equation}
  \label{eq:Heff}
  \tilde H = \sum_{k\sigma} (\epsilon_k - p_\sigma \tilde B)
  c^\dag_{k\sigma}c_{k\sigma} + \tilde U  \sum_i
  n_{i\uparrow}n_{i\downarrow}
\end{equation} 
where $\tilde U = \frac{J^2}{8 B}$, $\epsilon_k$ is the dispersion of
the original conduction electron band and $\tilde B = (B - J/4 -
\tilde U/2 )$ is the effective magnetic field. In this effective
model, the spin flip interaction between the conduction band and the
fully polarized $f$-spin background of the KLM, has been replaced by a
contact interaction between the $c$ electrons and the $f$ spins have
decoupled from the dynamics.

If the conduction electron band is particle hole symmetric, so that
$\epsilon_{k+Q} = - \epsilon_k$, the spin up hole and the spin down
electron Fermi surfaces of the effective model are perfectly nested.
Any non zero $\tilde U$ therefore induces magnetic ordering in the
plane perpendicular to the applied field. A mean-field decoupling,
with $\langle \vec s_i \rangle = ((-)^i m_x, 0, m_z)$, analogous to
the one performed in section~\ref{sec:meanfield}, yields the quasi
particle bands
\begin{equation}
  E_{\sigma,\pm}(k) = \pm \sqrt{
    (\epsilon_k - p_\sigma( \tilde B + \tilde U m_z) )^2 + (U m_x)^2 
  }
\end{equation}
and the mean-field equation determining $m_x$,
\begin{equation}
  \frac 2 {\tilde U} = \int_{-W}^0 \rho(\epsilon)d\epsilon \left[
    \frac 1 {E_{\uparrow,+}(\epsilon)} + \frac 1
    {E_{\downarrow,+}(\epsilon)} \right].
\end{equation}
In high magnetic fields, the up and down spin Fermi surfaces are well
approximated by circles of radii $W - \tilde B$ centered at
$(\pi,\pi)$ and $(0,0)$, respectively. We therefore can set
$\rho(\epsilon) = \rho_0 = v_F^{-1}$ to obtain
\begin{equation}
  \tilde U m_x \propto 2 (W - \tilde B - \tilde U m_z) 
  \exp\left( -\frac 1 {\rho_0 \tilde U} \right).
\end{equation}
The staggered magnetization and the quasi-particle gap are finite for
any finite $\tilde U$, as long as $\tilde B + \tilde U m_z < W$.  It
is easy to see that, $\tilde B + \tilde U m_z = W$ is just the
condition for system to fully polarize. This means that the staggered
magnetization vanishes only in the completely polarized system.
The completely polarized phase is a trivial insulator. The metallic
state is, therefore,
never obtained in the particle-hole symmetric case and is a bad
starting point for the perturbation expansion.

\subsection{Classical spins mean field}
\label{sec:classical-spins-mean}

We have seen that a large magnetic field suppresses both charge and
spin fluctuations on the $f$ sites.  The physics of the high-field
phase will, therefore, be well described by the KLM in which the $f$
spins are replaced by an array of statically arranged classical spins.
Let the spin configuration be
\begin{equation}
  \vec S_i = \frac 1 2 \left(
    \begin{array}[c]{c}
      \sin \theta \cos Q r_i \\
      -\sin \theta \sin Q r_i \\
      \cos \theta
    \end{array}
  \right).
\end{equation}
With $Q=(\pi,\pi)$, this corresponds to the same choice of $f$
magnetization as in section~\ref{sec:meanfield}, with $1/2 \sin \theta
= m_x$ and $1/2 \cos \theta = m_z$, so that the system is fully
polarized for $\theta = 0$. The problem is now reduced to one of the
non-interacting conduction electrons in an external magnetic field,
described by the Hamiltonian,
\begin{multline}
  \label{eq:clH}
  H = \sum_{k,\sigma} \left( \epsilon_k - p_\sigma \tilde B \right)
  c^\dag_{k\sigma} c_{k\sigma}  - N B \cos \theta \\
  + \frac{J \sin \theta}{4} \sum_k \left( c^\dag_{k-Q \uparrow} c_{k
      \downarrow} + c^\dag_{k+Q \downarrow} c_{k \uparrow} \right),
\end{multline}
with $\tilde B = B - J/4 \cos \theta$.  This is easily diagonalized to
find the quasi-particle bands ($k$ in the magnetic Brillouin zone),
\begin{equation}
  \label{eq:clbands}
  E_{\sigma,\pm}(k) =  
  \pm \sqrt{(\epsilon_k - p_\sigma \tilde B)^2 + (J/4)^2 \sin^2 \theta} .
\end{equation}
In the ground state the ``-'' bands are completely filled. Minimizing
the ground state energy and assuming the same circular Fermi surface
approximation as in the previous subsection, the mean field equation
determining the angle $\theta$ is obtained as,
\begin{multline}
  B \tan \theta = \rho_0 \int_{-W}^0 \sum_\sigma \left[ \left(\frac J
      4 \right)^2 \sin \theta
  \right.  \\
  \left.  -p_\sigma (\epsilon - p_\sigma \tilde B) \frac J 4 \tan
    \theta \right] \frac 1 {E_{\sigma,+}}
\end{multline}
For $\sin \theta \ll 1$ ($f$ moments almost aligned with the field) we
obtain,
\begin{equation}
  \label{eq:clmx}
  J m_x \propto \sqrt{(W-\tilde B)(W + \tilde B)} 
  \exp\left[-\frac{8B}{\rho_0 J^2} + \frac{4 \rho_0 B +8}{\rho_0 J} \right].
\end{equation}
The staggered magnetization vanishes only when $\tilde B = W$. It is
easy to see that this is exactly the condition for the system to fully
polarize. The full polarization field is equal to $W - J/4$ and agrees
with the one obtained using the effective Hamiltonian. The dominant,
small $J$ exponential dependence of the staggered magnetization $m_x
\propto \exp [ -8B/(\rho_0 J^2)]$ also agrees with the one obtained in
the previous section. The subleading $1/(\rho_0 J)$ correction to the
exponent arises because the Zeeman energy of the $f$ electrons has now
been taken into account. As the quasi-particle gap is proportional to
the staggered magnetization, the system stays insulating at all
fields.

\section{Quantum Monte Carlo}
\label{sec:qmc}

\begin{figure*}[tb]
  \centering \includegraphics[width=0.4\textwidth,height =
  0.2\textheight]{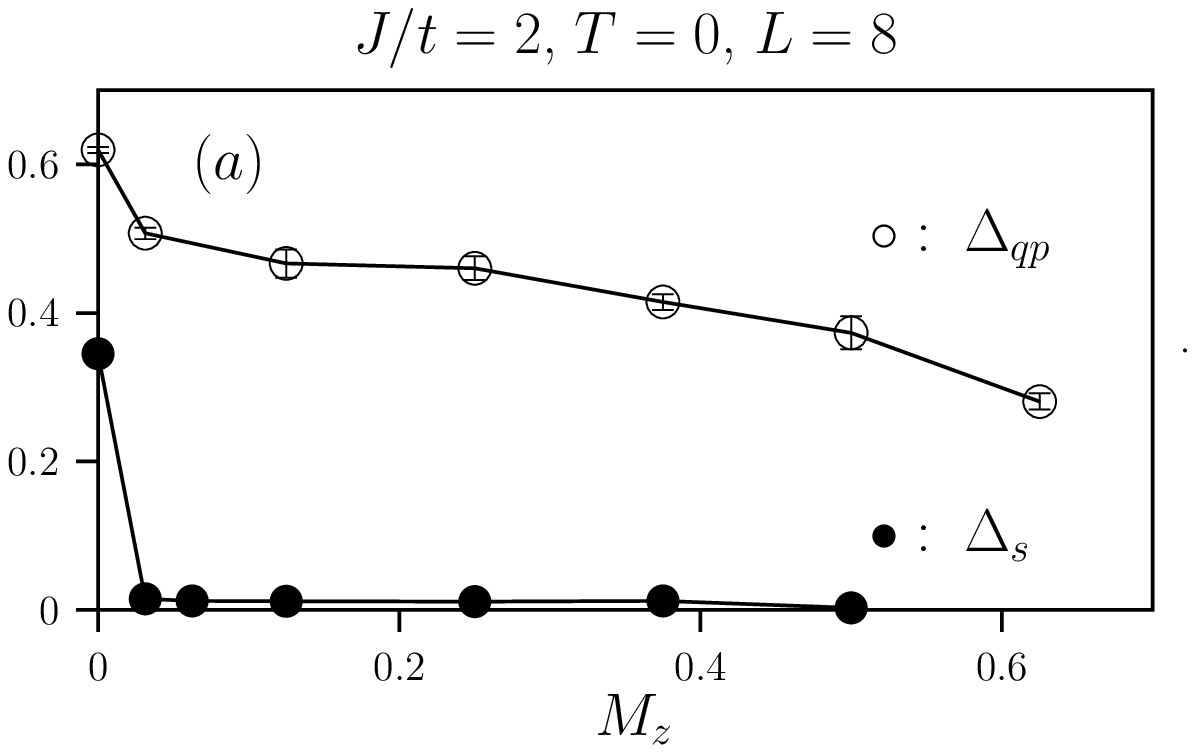}
  \includegraphics[width=0.4\textwidth,height =
  0.2\textheight]{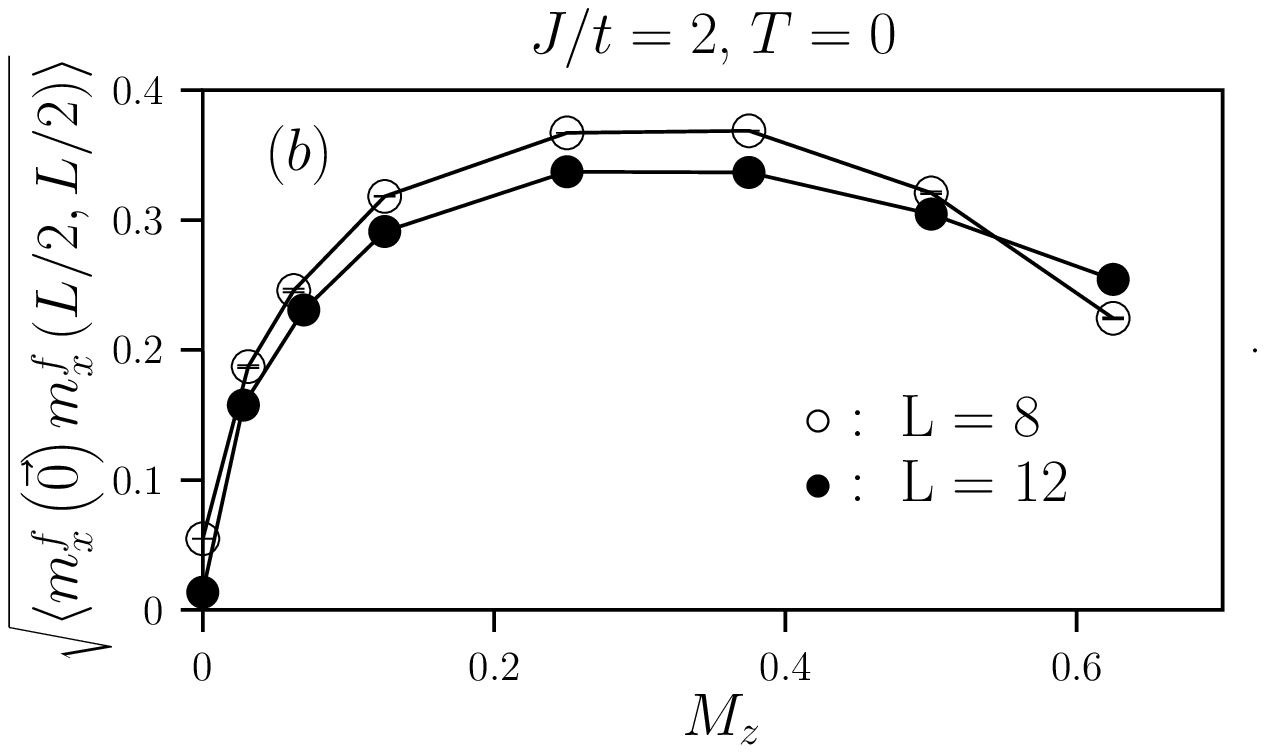} \\
  \includegraphics[width=0.4\textwidth,height =
  0.2\textheight]{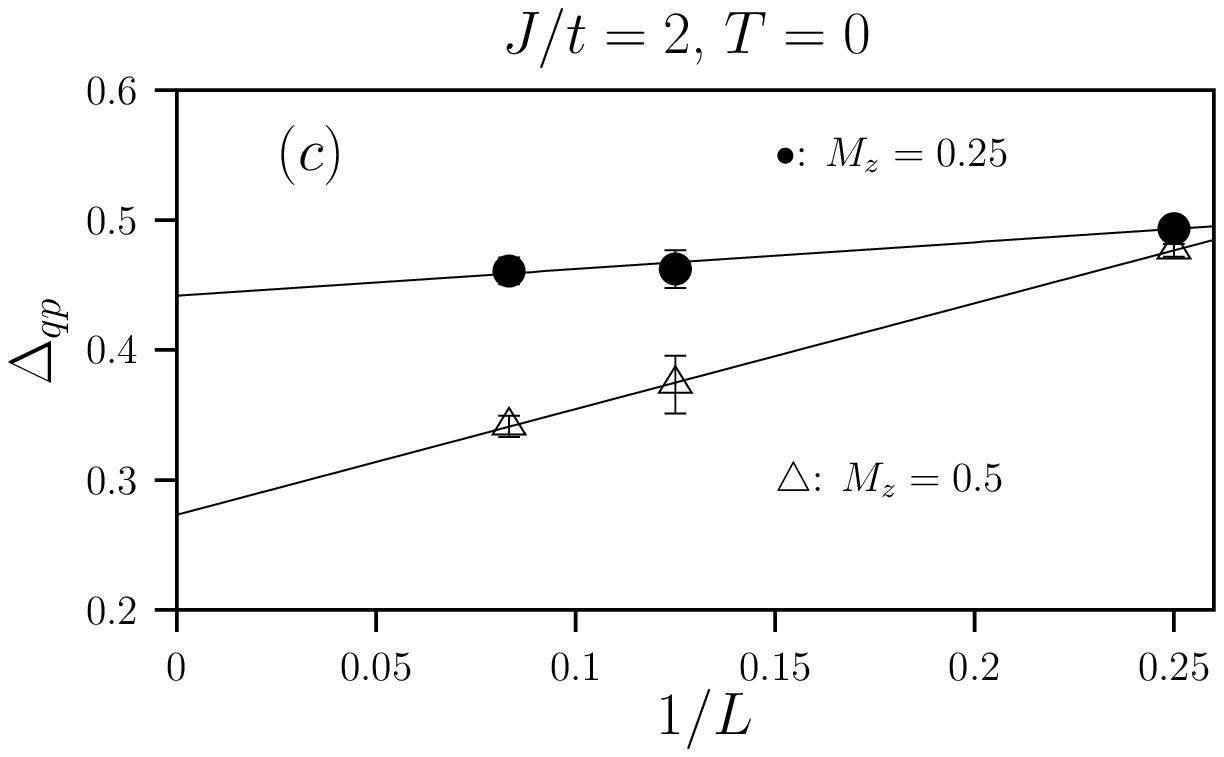}
  \includegraphics[width=0.4\textwidth,height =
  0.2\textheight]{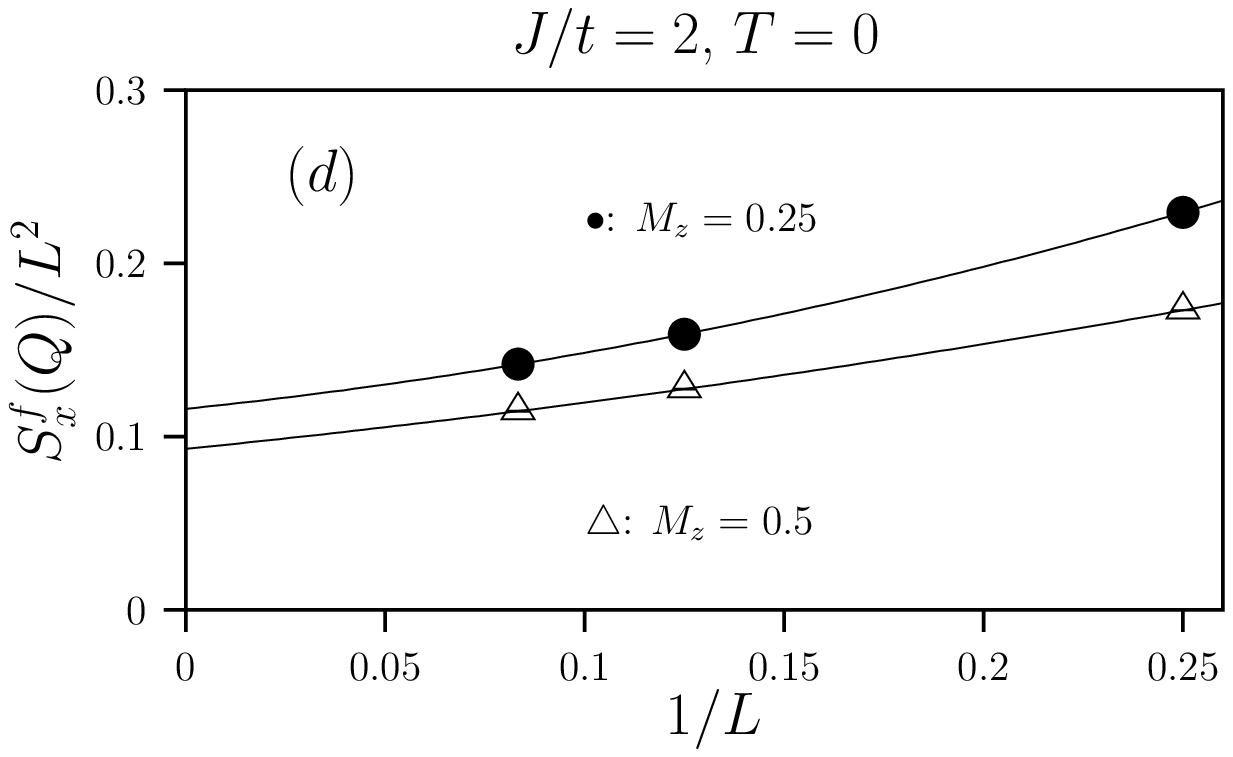}
\caption{(a) Quasi-particle and spin gaps as a function of the magnetization.  
  (b) x-component of the f-spin-spin correlations at the largest
  distance, $(L/2,L/2)$, on the $ L \times L$ lattice.  The Fourier
  transform of this quantity, $\sum_{\vec{r}} e^{i \vec{q} \cdot
    \vec{r} } \langle m_x^f(\vec{0}) m_x^f(\vec{r}) \rangle$, is
  peaked at the antiferromagnetic wave vector, $\vec{Q} = (\pi,\pi)$.
  (c) Size scaling of the quasi-particle gap at $M_z=0.25,0.5$. The
  data is consistent with a finite value of $\Delta_{qp}$ in the
  thermodynamic limit.  (d) Size scaling of $S_x^f(\vec{Q}=(\pi,\pi))
  = \sum_{\vec{r}} e^{i \vec{Q} \cdot \vec{r} } \langle m^f_x(\vec{r})
  m^f_x(\vec{0})\rangle$ for $M_z = 0.5, 0.25$ As apparent, the data
  is consistent with $S_x^f(\vec{Q}) \sim L^2$ thus signaling the
  presence of long-range antiferromagnetic order perpendicular to the
  applied field. }
\label{Gaps.fig}
\end{figure*}

In this section we present QMC simulations of the Kondo lattice model
in the magnetic field.  As in the zero field case, the sign problem
may be avoided only for particle-hole symmetric conduction bands.  To
compare with the mean-field results we adopt a projective approach in
which the ground state, $ | \Psi_0 \rangle $, is filtered out from a
trial wave function, $ | \Psi_T \rangle $, satisfying $ \langle \Psi_0
| \Psi_T \rangle \neq 0$.  In this algorithm, the ground state
expectation value of an observable $O$ is estimated via:
\begin{equation}
        \frac{ \langle  \Psi_0  | O  | \Psi_0 \rangle  }  {  \langle  \Psi_0  |  \Psi_0 \rangle } 
= \lim_{\Theta \rightarrow \infty}  
        \frac{ \langle  \Psi_T  | e^{- \Theta H /2 } O e^{ - \Theta  H / 2}  | \Psi_T \rangle  }  
             {  \langle  \Psi_T  | e^{-\Theta H} | \Psi_T \rangle }. 
\end{equation}
In the QMC, we evaluate the right hand side of the above expression at
finite values of $\Theta$ and then extrapolate to infinite values. The
details of the algorithm -- in particular the sign free formulation --
has been described extensively in Ref.~\onlinecite{capponi.2000}.
Since we are working in the canonical ensemble, the total
magnetization
\begin{equation}
        M_z = \frac{N^{\uparrow}_c  + N^{\uparrow}_f  - N^{\downarrow}_c   -  N^{\downarrow}_f  }{N_u}
\end{equation}
with $N_u$ the number of unit cells, is fixed during the simulations.

By measuring time displaced correlation functions, we can extract
quasi-particles as well as spin gaps.  Consider
\begin{multline}
  \langle \Psi_0 | S^-(-q,\tau) S^+(q,0) | \Psi_0 \rangle = \\
  \sum_n | \langle n | S^+(q) |\Psi_0 \rangle |^2
  e^{-\tau(E_n(q,N,S_z+1) - E_0(N,S_z)) }
\end{multline}
where $E_n(q,N,S_z)$ are eigenstates of $H$ with momentum $\vec{q}$,
particle number $N$ and total $z$-component of spin $S_z$.  From the
large $\tau$ behavior of the above correlation functions, we can
extract the energy difference $E_0(q,N,S_z+1) - E_0(N,S_z)$ from which
we can determine the spin gap:
\begin{equation}
        \Delta_{sp}(\vec{q})  = E_0(\vec{q},N,S_z+1) - E_0(N,S_z)  - h 
\end{equation}
where $ h = \left[ E_0(N,S_z + 1) - E_0(N,S_z -1) \right] / 2 $.  In
the same manner, we compute the quasi-particle gap from the
single-particle imaginary time displaced Green function.
\begin{equation} 
        \Delta_{qp}(\vec{k})  = E_0 ( \vec{k}, N+1,S_z)  - E_0(N,S_z) - \mu 
\end{equation}
with chemical potential: $\mu = [E_0(N+1,S_z) - E_0(N-1,S_z) ] / 2 $.
In Fig.~\ref{Gaps.fig} we plot the gaps as a function of total
magnetization at $J/t = 2.0$.

In the zero field case, the Kondo insulating state with finite
quasi-particle and spin gaps is realized.  At finite magnetizations
and according to the mean-field approach, we expect a canted
antiferromagnetic state and hence no spin gap.  Fig.~\ref{Gaps.fig}
confirms this point of view: the spin gap drops to zero at all finite
values of the magnetization within the accuracy of the numerical
simulation and the equal time spin-spin correlations in the plane
perpendicular to the magnetic field show long range antiferromagnetic
order. In the mean-field approach, magnetism stems from a Fermi
surface instability and due to perfect nesting opens a gap on all the
Fermi line. The QMC results of Fig.~\ref{Gaps.fig} are consistent with
this prediction, since, as apparent, the quasi-particle gap survives
at finite magnetizations.

\begin{figure*}[tb]
  \centering \includegraphics[width=\textwidth]{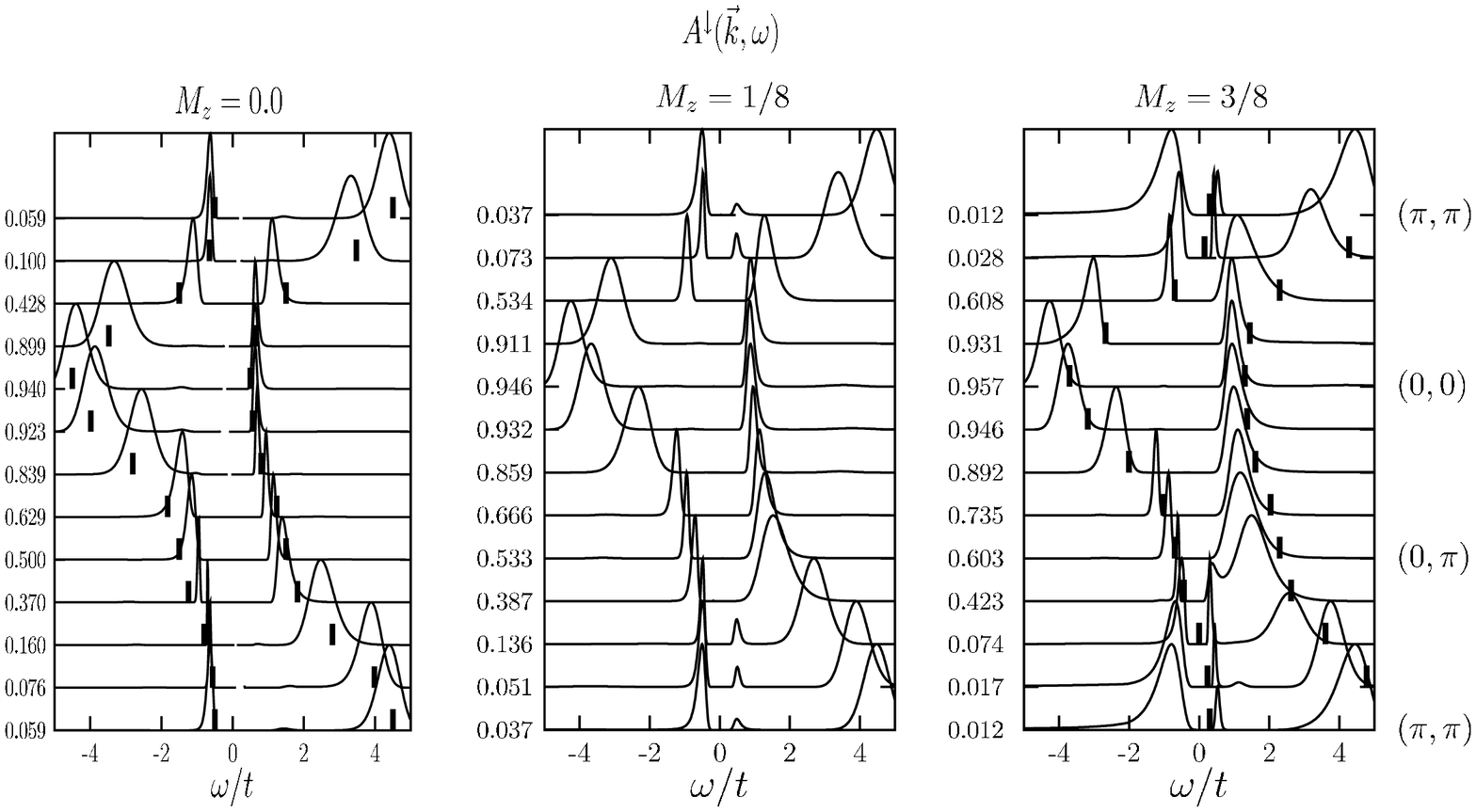}
\caption{Single particle spectral function in the down-spin sector as a function of magnetization 
  as obtained from analytical continuation of the QMC results with the
  Maximum Entropy method.  The solid vertical lines correspond to the
  pole position of hybridized bands (see text). On the left hand side
  of each Fig. we have listed the single particle occupation number
  $n_{ \vec{k},\downarrow} $ from which one can determine the weight
  under the spectral function (see Eq. \ref{Sum_rule}).}
  \label{akom.fig}
\end{figure*}

To further compare the mean-field approach to the QMC we have used the
Maximum Entropy method to obtain the single particle spectral function
$A(\vec{k},\omega)$.  At zero field in the Kondo insulating phases,
the dominant features of the spectral function are well described by
hybridized bands (solid vertical lines in Fig. \ref{akom.fig}a). From
the single particle occupation number $ n_{\vec{k},\sigma} = \langle
c^{\dagger}_{\vec{k},\sigma} c_{\vec{k},\sigma} \rangle $ listed on
the left hand side of Figs \ref{akom.fig}a-c we can extract the total
weight under the ``photoemission'' ($\omega < 0$) and the ``inverse
photoemission'' ($\omega < 0$) spectra since
\begin{equation}
\label{Sum_rule}
        \int_{-\infty}^{0} {\rm d } \omega A^{\sigma}(\vec{k},\omega) = \pi n_{\vec{k},\sigma}  \;\;
        \int_{0}^{\infty} {\rm d } \omega A^{\sigma}(\vec{k},\omega) = \pi(1 - n_{\vec{k},\sigma} ).
\end{equation}
In particular, one sees that the photoemission (inverse
photo-emission) spectrum in the vicinity of $ \vec{k} = (\pi,\pi) $ $
(\vec{k} = (0,0)) $ has a small weight. Those heavy bands stem from
the Kondo screening.

In the mean field approach, the effect of a magnetic field is to shift
the spin down band up in energy until it ultimately crosses the Fermi
surface, thus generating a metallic state. This metallic state is
however unstable due to the underlying particle-hole symmetry.  In
Fig.~\ref{akom.fig}c we compare the finite field results with a rigid
shift of the hybridized bands. The data are compatible with the
interpretation that the down spin band has indeed crossed the Fermi
surface but that at the crossing point the magnetic instability opens
a gap.  Furthermore on the photo-emission side around the $(\pi,\pi)$
point we see a very weak feature which we can identify as a shadow of
the up band which has dropped below the Fermi energy in the vicinity
of the $\vec{k} = (0,0) $ point.

Breaking of the spin symmetry by the magnetic field suppresses Kondo
screening.  We hence expect the weight of the features in the spectral
function stemming from Kondo screening to be suppressed as a function
of growing magnetization.  For example, consider the inverse
photoemission at $\vec{k} = (0,0) $ in Fig. \ref{akom.fig}.  As
apparent the weight of this feature is reduced as a function of
growing values of $M_z$ and will vanish when the $f$-spins become
fully polarized. In our simulations, the $f$-spins are never
completely aligned with the field as long as the system is not
completely polarized, i.e. as long as $N_c^\uparrow + N_f^\uparrow <
N_u$. This supports the conclusion that the metallic state is never
induced by the field.



\section{Conclusion}
\label{sec:conclusion}

We studied the magnetic field induced quantum phase transition in the
2D particle-hole symmetric Kondo insulators using: i) a mean field
approximation appropriate in the small $U$ limit of the periodic
Anderson model, ii) two mean field approximations appropriate in the large
field limit of the Kondo lattice model and iii) a quantum Monte Carlo
simulation of the particle-hole symmetric 2D Kondo lattice model in
the field.

We find a
magnetic field induced quantum phase transition from a paramagnetic
insulator into a canted antiferromagnetic insulator ground state. In
the particle-hole symmetric case we studied, the antiferromagnetism
can be understood as a spin density wave type instability of the
perfectly nested quasi-particle Fermi surfaces that would arise
in the field in the absence
of interactions. Because of the perfect nesting, any finite
interaction is a relevant perturbation and results in a
finite quasi-particle gap. 
Consequently, the ground state of the interacting system remains
insulating in all fields.

We conclude that the recently proposed insulator to metal transition
induced by the field\cite{beach.03} is likely to be an artifact of the
large-$N$
approximation to the Kondo lattice model in the particle-hole
symmetric case. If,
however, the particle-hole symmetry is violated a field-induced
metal-insulator transition is possible in certain parameter ranges.

We find that the qualitative features of the phase diagram as well as
of the quasi-particle excitations are well described by a simple mean
field approximation to the periodic Anderson model. The magnetic field
explicitly breaks the spin rotation symmetry and suppresses the charge
fluctuations on $f$ electrons, essentially by fully polarizing the $f$
band.

The band structure of the Kondo insulators is non-trivial and
deviations from particle-hole symmetry are to be expected in the real materials.
In the absence of perfect nesting there would be a critical value of
the field, controlled essentially by the nesting mismatch, at which
the gap will close on some parts of the Fermi surface. Therefore one
expects to eventually find a metallic state induced by the field.
Finally we would like to mention that the conclusions we draw here are
valid also for the three-dimensional systems, where the same kind of
nesting features would appear for perfect particle-hole symmetry.

{\bf Acknowledgments} We would like to thank P.A. Lee, T.M. Rice, K.D.S. Beach and S. Wehrli
for helpful discussions. This work was financially supported by the
Swiss Nationalfonds and, in particular, by the NCCR program MaNEP.
F.F. Assaad thanks the DFG for financial support under the grant
number of AS 120/1-1 as well as the hospitality of the ITP of the
ETHZ where part of this work was carried out.  The calculations were
performed on the Cray-T3E as well as on the IBM-p690 in J\"ulich. We
thank this institution for generous allocation of CPU time.

\appendix

\section{Mean field decoupling of the Hubbard term in the canted anti-ferromagnetic phase}
\label{sec:app-meanfield}

We want to decouple the interaction term in the PAM, in the presence
of a canted staggered magnetization. To this end, we select the spin
quantization axis 
at each site to point in the direction of the local magnetization,
$\vec m_i$. Using the
operator identity $ f^\dag_{i\uparrow_i} f_{i\uparrow_i}
f^\dag_{i\downarrow_i} f_{i\downarrow_i} = 1/4 (f^\dag_{i\uparrow_i}
f_{i\uparrow_i} + f^\dag_{i\downarrow_i} f_{i\downarrow_i})^2 - 1/4
(f^\dag_{i\uparrow_i}f_{i\uparrow_i} - f^\dag_{i\downarrow_i}
f_{i\downarrow_i})^2$, where $\uparrow_i (\downarrow_i)$ denotes the
spin with respect to the local quantization axis, the interaction term
can be decoupled as
\begin{multline}
  \label{eq:decoupling}
  U \sum_i 
  f^\dag_{i\uparrow_i} f_{i\uparrow_i} f^\dag_{i\downarrow_i} f_{i\downarrow_i}
  =
  \frac {Un_f}2 \sum_i ( f^\dag_{i\uparrow_i} f_{i\uparrow_i} 
  + f^\dag_{i\downarrow_i} f_{i\downarrow_i} ) \\
  - N \frac{Un_f^2} 4
  - U \sum_i |\vec m_i| (  f^\dag_{i\uparrow_i} f_{i\uparrow_i} 
      - f^\dag_{i\downarrow_i} f_{i\downarrow_i} ) + U |\vec M|^2 \\
   + \text{Fluct.} 
\end{multline}
where $n_f = \langle f^\dag_{i\uparrow} f_{i\uparrow} +
f^\dag_{i\downarrow} f_{i\downarrow} \rangle$ is the average occupancy
of the $f$ site and ``Fluct.'' denotes the terms neglected in the
mean-field approximation. After the decoupling, a spin axis rotation to a
common quantization axis (given by the direction of the external
field) is performed using $\hat R = \exp \left( \sum_i -i/2 \vec \theta_i
\cdot \vec S^f_i \right)$ with $\vec \theta_i$ being the vector pointing
along $\vec B \times \vec m_i$ and of magnitude equal to the angle
between $\vec B$ and $\vec m_i$. Since,
\begin{equation}
  |\vec m_i| \hat R \tfrac 1 2 (  f^\dag_{i\uparrow_i} f_{i\uparrow_i} 
      - f^\dag_{i\downarrow_i} f_{i\downarrow_i} ) \hat R^\dag = \vec m
        \cdot \vec S^f_i, 
\end{equation}
this yields the mean field Hamiltonian,
Eq..\ref{eq:HMF} of section~\ref{sec:meanfield}. 

To obtain the behavior of the staggered magnetization in fields $B >
\Delta_{qp}^0$ and at small $U$ we need to solve the mean field
equations~\ref{eq:mf}, in the limit when $U m_x \rightarrow 0$. As the only $k$
dependence of the quasi-particle bands comes through the $k$
dependence of the conduction electron energy, the
summations over $k$ are readily transformed into integrals over the
conduction electron energy, thus yielding ($\rho(\epsilon)$ is the
conduction electron DOS),
\begin{equation}
  \frac{\partial}{\partial (Um_x)} \left[ 
  \int_{-W}^0 \rho(\epsilon) 
  d \epsilon (E_{h,\pm}^\uparrow(\epsilon) +
  E_{h,\pm}^\downarrow(\epsilon)) \right]
  + 2 \frac {Um_x}{U} = 0.
\end{equation}
For $Um_x \rightarrow 0$, the dominant contribution to the integral
comes from the band crossing the Fermi surface at $\epsilon_0$ (given
by Eq.~\ref{eq:epsilon0}). For
$\Delta_{qp}^0 < B < V$, this is $E_{h,+}^\downarrow$. For small $U$ and
close to $\epsilon_0$ we can write,
\begin{equation}
  E_{h,+}^{\downarrow}(\epsilon) = \sqrt{\alpha^2 (Um_x)^2 + \beta^2
  (\epsilon - \epsilon_0)^2}, 
\end{equation}
with
\begin{align}
  \alpha &= \frac{V^2}{(B+Um_z)^2 + V^2} \\
  \beta &= \frac{(B+Um_z)^2}{(B+Um_z)^2 + V^2}.
\end{align}
The logarithmically divergent part of the mean field equation can now
be written as (neglecting the
non-divergent contributions),
\begin{equation}
  \frac 1 U = \frac{\alpha^2}{\beta} \int_0^{\epsilon_c}
  \frac {d\epsilon' \rho(\epsilon_0+\epsilon')}
  {\sqrt{\left(\frac{\alpha U m_x}{\beta}\right)^2 + \epsilon'^2}},
\end{equation}
where we have
introduced a cutoff $\epsilon_c$ which does not influence the
exponential dependence.
Eq.~\ref{eq:mx} in the text now follows by elementary integration.

\section{Effective Hamiltonian in the large field}
\label{sec:app-largeB}

The ground state of the $J=0$ KLM in the magnetic field is
non-degenerate. All the $f$ spins are polarized in the direction of
the field. The $f$ spin flip is an excitation with an energy gap
of the size of the Zeeman energy.
Let $P_n$ denote the portion of the Hilbert space with $n$ localized
spins pointing opposite to the magnetic field and let $\mathcal P_n$
be the corresponding projector.
In large magnetic fields, $B \gg J$, one expects the ground state and
the low lying excitations to lie dominantly in $P_0$ and have only
small components in the $P_n$, with $n>0$, subspaces.
We split the $H_{KLM}$ into a $P_n$ diagonal part 
\begin{multline}
  \label{eq:H0}
  H_0 = \sum_{k\sigma} (\epsilon_{k\sigma} - p_\sigma B ) 
  c^\dag_{k\sigma} c_{k\sigma}
  - 2 B \sum_\sigma p_\sigma S_i^z \\
  + \frac{J}{2} \sum_i S_i^z ( c^\dag_{i\uparrow}
    c_{i\uparrow} - c^\dag_{i\downarrow} c_{i\downarrow})
\end{multline}
and the spin flip part
\begin{equation}
  V = \frac J 2 \sum_i \left( c^\dag_{i\uparrow} c_{i\downarrow} 
    S_i^- +
  c^\dag_{i\downarrow} c_{i\uparrow} S_i^+ \right) .
\end{equation}
Let $S$ be a Hermitian operator, such that
\begin{equation}
  \label{eq:S1}
  [H_0, S] = - V .
\end{equation}
The effective Hamiltonian, obtained by applying the canonical transformation
\begin{equation}
  \tilde H = e^{-S} H_{KLM} e^S
\end{equation}
has no matrix elements between the states in $P_0$ and $P_{n>0}$ of
order less then $\mathcal O ( J(J/B)^2 )$.
We can therefore obtain the low-energy dynamics of the original problem
correctly to order $J^2/B)$ by considering only the $P_0$
part of the Hilbert space and the Hamiltonian,
\begin{equation}
  \label{eq:HeffP0}
  \tilde H^{P_0} = \mathcal P_0 e^S H_{KLM} e^{-S} P_0.
\end{equation}
By expanding and rearranging the exponentials one obtains
\begin{equation}
  \label{eq:commutators}
  \tilde H^{P_0} = \mathcal P_0 \left( H_0 + \frac 1 2 \left[V,
  S\right] + \cdots \right) \mathcal P_0
\end{equation}
Using Eq.(\ref{eq:S1}) it is easy to obtain the matrix elements
of $S$ between the eigenstates of $H_0$ from which the operator form
easily follows,
\begin{equation}
  \label{eq:S}
  S = \frac{J}{4B} \frac 1 {\sqrt N} \sum_{kq} \left(
    S^+_q c^\dag_{k-q\downarrow} c_{k\uparrow} 
    - S^-_q c^\dag_{k-q\uparrow} c_{k\downarrow} \right),
\end{equation}
thus yielding,
\begin{equation}
  [V,S] = \frac {J^2}{4B} 
  \frac 1 N \sum_{qkk'} \left[ S_q^- c^\dag_{k\uparrow}c_{k-q\downarrow},
    S_q^+ c^\dag_{k'\downarrow}c_{k'-q\uparrow} \right] .
\end{equation}
Evaluating the commutator and projecting on the $P_0$ subspace, yields
\begin{equation}
  \mathcal P_0 [V,S] \mathcal P_0 = 
  - \frac {J^2}{4B} \frac 1 N \sum_{kk'q} 
  c^\dag_{k'\downarrow}c_{k'-q\uparrow} c^\dag_{k\uparrow}c_{k-q\downarrow}.
\end{equation}
Substituting into Eq.~(\ref{eq:commutators}) and neglecting the terms
corresponding to the dots, which are of order $\mathcal O(J^3/B^2)$
one obtains the effective Hamiltonian,
\begin{equation}
  \tilde H = H_0
  + \frac{J^2}{8B}\sum_i s^z_i 
  + \frac{J^2}{8B} \sum_i n_{i\uparrow} n_{i\downarrow}
   - \frac{J^2 N}{16 B}.
\end{equation}
The last term in the above equation is a constant and can be
dropped, the $s^z_i$ term is the contribution to the effective uniform
magnetic field.
This completes the derivation of the effective model
described in section~\ref{sec:largeB}.

\bibliography{references}

\end{document}